\documentclass[9pt,twocolumn,twoside]{osajnl}
\usepackage[nolist]{acronym}
\usepackage{float}
\usepackage{multirow}
\journal{jocn} 

\definecolor{mycol}{rgb}{0.49, 0.18, 0.56}

\setboolean{shortarticle}{false}

\title{Two-tier PON virtualisation with scheduler synchronization supporting application-level ultra-low latency in MEC based Cloud-RAN, using MESH-PON}

\author[1,*]{Sandip Das}
\author[1]{Frank Slyne}
\author[1]{Daniel Kilper}
\author[1]{Marco Ruffini}

\affil[1]{The University of Dublin, Trinity College, CONNECT research centre.}

\affil[*]{Corresponding author: dassa@tcd.ie}

\begin{abstract}
	Ultra-low end-to-end latency is one of the most important requirements in 5G networks and beyond to support latency-critical applications. Cloud-RAN and \ac{MEC} are considered as the key driving technology that can help  reduce end-to-end latency. However, the use of MEC nodes poses radical changes to the access network architecture. As it brings the processing and the networking services closer to the edge, it often requires network functions (for example, the CU/DU stack and the application processing) to be distributed across different MEC sites. Therefore, a novel transport mechanism is needed to efficiently coordinate and connect network functions across MEC nodes.
	
	In order to address this challenge, we propose a novel two-tier virtualized PON transport method with schedulers coordination over a virtualised and sliced MESH-PON architecture. While a MESH-PON architecture enables direct communication between MEC nodes that are hosting CU/DU and/or the application processing, our method provides a two tier virtualised PON transport scheme with coordinated schedulers. This approach greatly reduces latency incurred in transporting signals across the different PON tiers, while maintaining the flexibility of the multi-tier methods. We show that our proposed scheme can achieve end-to-end application-level latency below 1ms or 2ms, depending on the network configurations.
	
\end{abstract}

\setboolean{displaycopyright}{true}

\begin{document}
	
	\maketitle
	
\begin{acronym}
	\acro{AR}{Augmented Reality}
	\acro{QoS}{Quality of Service}
	\acro{C-RAN}{Cloud Radio Access Networks}
	\acro{CGS}{Coordinated Grant Scheduling}
	\acro{CU}{Central Unit}
	\acro{CTI}{Coordinated Transport Interface}
	\acro{FBG}{Fiber Bragg Grating}
	\acro{GF}{Grant Free}
	\acro{ITS}{Intelligent Transport Systems}
	\acro{MAIO}{Mixed Analytical Iterative Optimization}
	\acro{MFH}{Mobile Fronthaul}
	\acro{RU}{Radio Unit}
	\acro{BBU}{Baseband Unit}
	\acro{DU}{Distributed Unit}
	\acro{DU/CU}{Distributed Unit/Centralised Unit}
	\acro{DCI}{Downlink Control Information}
	\acro{NR}{New Radio}
	\acro{PON}{Passive Optical Network}
	\acro{vPON}{virtual-PON}
	\acro{ODN}{Optical Distribution Network}
	\acro{TWDM}{Time-Wavelength Division Multiplexing}
	\acro{DBA}{Dynamic Bandwidth Allocation}
	\acro{MEC}{Multi Access Edge Computing}
	\acro{CO}{Central Office}
	\acro{OLT}{Optical Line Terminal}
	\acro{RV}{Random Variable}
	\acro{GC}{Grant Cycle}
	\acro{ONU}{Optical Networking Unit}
	\acro{PLOAM}{Physical Layer Operation and Maintenance}
	\acro{PRB}{Physical Resource Block}
	\acro{eCPRI}{evolved Common Public Radio Interface}
	\acro{BS}{Base Station}
	\acro{SPS}{Semi-Persistent Scheduling}
	\acro{TDMA}{Time Division Multiple Access}
	\acro{TTI}{Transmit Time Interval}
	\acro{VRF}{Variable Rate Fronthaul}
	\acro{vPON}{Virtualized PON}
	\acro{UE}{User Equipment}
	\acro{LLS}{Low Layer Split}
	\acro{vRAN}{Virtualized RAN}
	\acro{WLB}{Wavelength Loop Back}
	\acro{WPF}{Wavelength Pass Filter}
\end{acronym}

	\section{Introduction}
	As the the deployment of 5G networks has moved past its initial phase, operators are pushing forward to find technological solutions to fine tune their fronthaul/backhaul networks to address key requirements for 5G and beyond. Among these, the support for ultra-low latency applications (i.e., of the order of 1ms) is important to enable mission-critical applications such as \ac{ITS}, industry 4.0, public safety, including use of \ac{AR} technology \cite{5G-NGMN-Verticals}. \ac{C-RAN}, and \ac{MEC} is rapidly replacing the legacy architecture as it can better support network densification and local data processing and storage. From a networking perspective, when considering end-to-end (i.e., from source to destination at the application level) we could break down the network elements contributing to overall latency into three sections: the RAN (due to wireless resource scheduling, transmission distance and stack processing), the fronthaul transmission from \ac{RU} to \ac{DU/CU} (due to transmission distance and data scheduling if operating over a PON) and the transport from the DU/CU towards the MEC node running the application. 
	
	In a traditional RAN, the access latency is the amount of time the \ac{UE} application traffic needs to wait for allocation of uplink resources before transmission, which is generally assigned to the UE via a set of \ac{DCI} messages in 5G \ac{NR}. The latency between buffer status report by the UE and the corresponding uplink resource grant allocation via DCI messages is 4 time slots (e.g., 2 ms for 0.5 ms slot duration \cite{5G-NR_Dahlman-Scheduling}). This is the largest contributing factor to RAN access latency. In order to address this bottleneck, \ac{CGS} (for uplink) and \ac{SPS} (in downlink) were proposed \cite{5G-NR_Dahlman}, which semi-statically pre-allocate uplink resources (typically a group of \acp{PRB}) to UEs so that they can send their uplink traffic without making a request, thus avoiding waiting for the scheduling of uplink resources. 
	
	The second source of latency, as mentioned above, occurs due to the uplink scheduling of fronthaul, when the RAN is transported over a PON. This latency becomes critical if the RAN is employing a functional split that is a split 6 (between MAC and PHY) or above \cite{ORAN_FH_Standard}.  
	Here, if the PON and RAN schedulers are not coordinated properly, data from the UE will need to queue at the ONU side waiting for the PON upstream grant to be provided by the OLT. This coordination issue was recently solved with the development of cooperative Dynamic Bandwidth Allocation (Co-DBA) \cite{Co-DBA} implemented over a \ac{CTI}\cite{O-RAN-CTI}. This requires the OLT to fetch prior UE uplink scheduling information from the DU and use it to estimate the fronthaul packet size and arrival time at the ONU that is connected to the RU. Based on this information, the OLT can pre-assign uplink resources to the ONU so that the packets from the UEs undergo minimal queuing once they arrive at the ONU from RU. 
	
	However, this does not solve the issue of latency at the application level. As mentioned above, low latency requires the use of methods like \ac{CGS} in the RAN, where information about incoming data from the UE is not known in advance and thus cannot be passed to the OLT for CTI coordination. 
	CTI currently does not support a RAN that uses the CGS for ultra-low latency. In this work we propose an updated CTI that can support low-latency RAN operations, thus addressing this issue. 
	
	The third source of network latency is the data transmission between the DU/CU and the server running the application. Typically, this is sent over the network to a \ac{CO} or a \ac{MEC} node, and can involve multiple layer 2 or layer 3 hops, depending on the network configuration. A second contribution of this work is to address these shortcomings through the a new MESH-PON approach, described below. 
	\begin{figure*}[h]
		\centering
		\includegraphics[clip, trim={0, 0, 0, 0}, width=\linewidth]{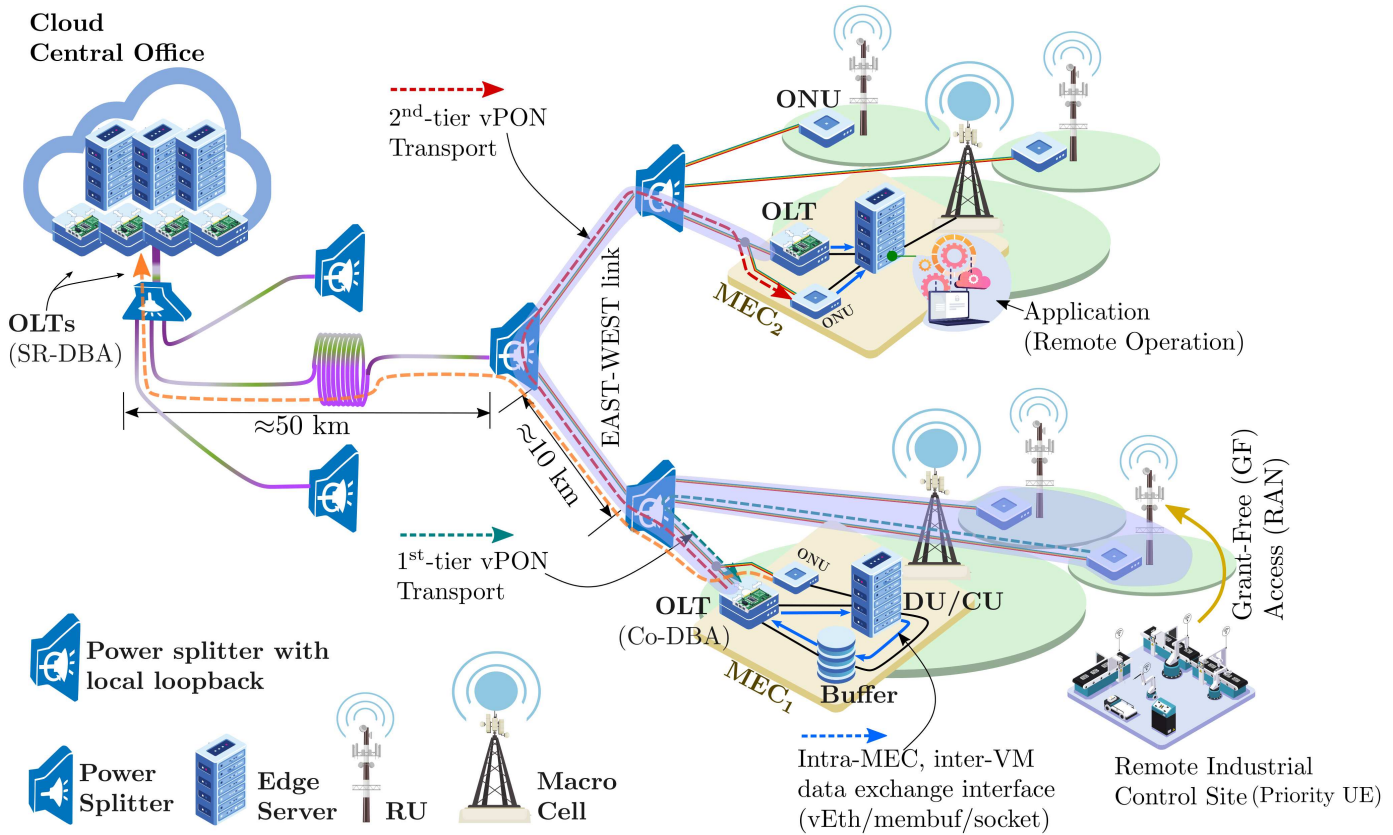}
		\caption{System architecture and the use case}
		\label{fig:SystemArc}
	\end{figure*}
	
	We base our architecture on virtualised PONs (vPONs) \cite{vDBA} operating over a mesh access topology \cite{MESH-Networks}. A MESH-PON makes use of technology such as wavelength reflectors at splitter locations as in \cite{MESH-JOCN} (or other configurations as in \cite{Wong,Pfeiffer}) to enable direct communications between end points (i.e., without the need of OEO conversion and packet scheduling by the OLT located at the source of the PON tree in the central office). It should be noticed that a legacy PON that does not support mesh connectivity can only operate as a point to multipoint. Thus the only option to achieve connectivity between end points is to communicate to an OLT located at the source of the PON tree (i.e., at the CO), which accumulates considerable latency over each round trip. In our MESH-PON, a vPON can be dynamically created among a set of nodes that require direct communications. For example a number of small cell RUs can create a vPON that includes an MEC node that hosts the DU/CU servers controlling the small cell RUs. The virtualisation aspect (combined with a flexible and tunable physical layer) enables the connectivity among this set of nodes to be created and modified dynamically (i.e., if due to a change in load or services a set of small cell RUs needs to connect to a different MEC node). In our previous work, \cite{ECOC2022}, we presented a method for coordinating transmission from the RAN to a first MEC node that hosts the DU/CU (first tier) and then from there to another MEC node (second tier) that hosts the application. 
	It is worth noting that it is possible for the same MEC node to host both DU/CU and applications. However, our solution allows for a more general case where a functional chain may be spread across multiple locations. This can for example support multi-tenancy \cite{C-RAN-multitenancy}: the owner of the C-RAN and that of the application can be different entities that run their services from different MEC nodes. 
	In this work, we extend our work in \cite{ECOC2022} in two ways. Firstly, we provide a detailed communication protocol and the management of the vPON slices to facilitate the schedulers synchronization for achieving application-level end-to-end low-latency, and analyze the latencies involved in the various stages of the overall end-to-end transport. Secondly, we examine the impact of various network factors (such as traffic load at the RU and average OLT downlink load in the second-tier transport) on the end-to-end latency of applications in our proposed architecture, and discuss potential solutions. Therefore, the overall contributions of this work can be summarised as follows:
	\begin{enumerate}
		\item In the first tier, we propose an enhanced cooperative DBA which can support CGS, to achieve ultra-low latency both in the RAN and fronthaul PON transport.
		\item In the second tier, we propose an uplink-to-downlink switching mechanism between virtual PONs that minimizes latency towards the application MEC node.
		\item We have provided an understanding of how different network parameters, such as traffic load at the RU and average OLT downlink load in the second-tier transport, impact the end-to-end latency of the proposed architecture, and possible strategies to overcome any shortcomings
	\end{enumerate}
	
	The rest of this article is organized as follows: In Section \ref{sec:SystemModel}, we provide the system architecture of our proposed two-tier vPON transport method. Here we also discuss the details of transport protocol and coordination between two schedulers ($1^\text{st}$ and $2^\text{nd}$ tier). In Section \ref{sec:Simulation}, we provide details of the discrete event simulation to carry out performance evaluation, whose results are then provided in Section \ref{sec:Results}. Finally, we conclude this article in Section \ref{sec:Conclusion}.
	
	\section{System Architecture} \label{sec:SystemModel}
	Fig. \ref{fig:SystemArc} presents the proposed system architecture and use case. We consider a fixed-mobile converged access scenario, where a mesh TWDM-PON similar to our proposed architecture in \cite{MESH-Networks} is used for sharing C-RAN fronthaul with residential broadband users (not shown in the figure). We also consider MEC nodes hosted at the macro cell site for providing low-latency RAN and application processing. For the low-latency RAN, we target types of URLLC applications with latency requirements of the order of 1ms \cite{5G-NGMN-Verticals}, for example the scenario of real-time control application of remote industrial site as shown in Fig. \ref{fig:SystemArc}. In order to meet this tight end-to-end latency, we propose the following coordinated two-tier vPON scheduling method.
	
	The proposed architecture features a number of PON endpoints that serve small cell RU sites, which are equipped with wavelength-tunable ONU capability, and MEC nodes that have tunable OLT capability. This enables the OLTs at the MEC nodes to simultaneously communicate with multiple RUs and other OLTs at other MEC nodes via EAST-WEST PON connectivity (illustrated with red-colored dashed-line in Fig. \ref{fig:SystemArc}). Such direct connectivity between MEC nodes and multiple RUs is achieved by reflecting back selected wavelengths through Fiber Bragg Gratings (FBGs) at the level-1 splitter locations which is shown as "Splitter with local loopback" in Fig. \ref{fig:SystemArc}. Residential users (not shown here) can be served by regular, low-cost, non-tunable ONUs and connect to the central office via NORTH-SOUTH, point-to-multipoint connectivity. 
	In order to transport fronthaul data with low latency, ONUs connected to RUs providing URLLC services (referred to as priority-UE traffic) can create a virtual PON slice with the OLT located at nearby MEC nodes (MEC-1 in this case). We refer to this as the $1^{st}$-tier and its path is illustrated with the green-colored dashed line in Fig. \ref{fig:SystemArc}. Finally, similar to our MESH-PON architecture in \cite{MESH-JOCN}, the tunable ONUs at the MEC sites provide a connection to the central office, ensuring that a communication channel is always available via NORTH-SOUTH connectivity (illustrated in orange-colored dashed line) for exchanging control information, such as receiving vPON slice configurations from the central office. To fully understand the practical feasibility of the MESH-PON connectivity, including the splitter architecture, wavelength planning, physical lightpath creation via vPON formation, and power budget analysis, we encourage readers to refer to our previous work in \cite{MESH-JOCN} and \cite{MESH-Networks} for a more in-depth examination. 
	
	\begin{enumerate}
		\item \textsl{\Large Communication Protocol:}
		
		In order to achieve low end-to-end latency in fronthaul, the first tier requires a novel, enhanced Co-DBA mechanism, to efficiently incorporate the CGS resources that can deliver URLLC. The conventional Co-DBA \cite{Co-DBA} utilizes mobile scheduling information 4-TTI (4 NR slot time for the 5G case) in advance from the CU/DU to derive the uplink grants for efficient fronthaul transport with ultra-low latency. The O-RAN standard for Cooperative Transport Interface \cite{O-RAN-CTI} outlines the interface definition and messaging protocol between the CU/DU and OLT for achieving this Co-DBA. However, the conventional Co-DBA relies on the fact that the proper mobile scheduling information is available 4-NR slots prior and does not take into account the traffic scheduling at CGS resources, where information is not typically known in advance. To encounter this, our proposed enhancement to the conventional Co-DBA incorporates information about the semi-static allocation of CGS resources from RRC to accurately calculate DBA and account for traffic at the CGS resources. As the Radio Resource Control (RRC) block in the CU semi-statically allocates a set of CGS resources to a specific UE for URLLC services, this can be made available from the CU/DU via CTI interface, and our algorithm passes this information to the OLT, so that it can calculate uplink grants using our proposed enhanced Co-DBA considering the allocation of CGS resource particular to the RU.
		A conservative approach, implemented in this work, is to consider the allocated CGS resources regardless of how many PRBs the UE is actually using. However, efficiency could be further improved by measuring the current UE traffic and then estimate the percentage of the CGS resources that are occupied in the uplink and use it along with the typical mac scheduling information for calculating the grants.
		
		\begin{figure*}[h]
			\includegraphics[clip, trim={0, 0, 0, 0}, width=\linewidth]{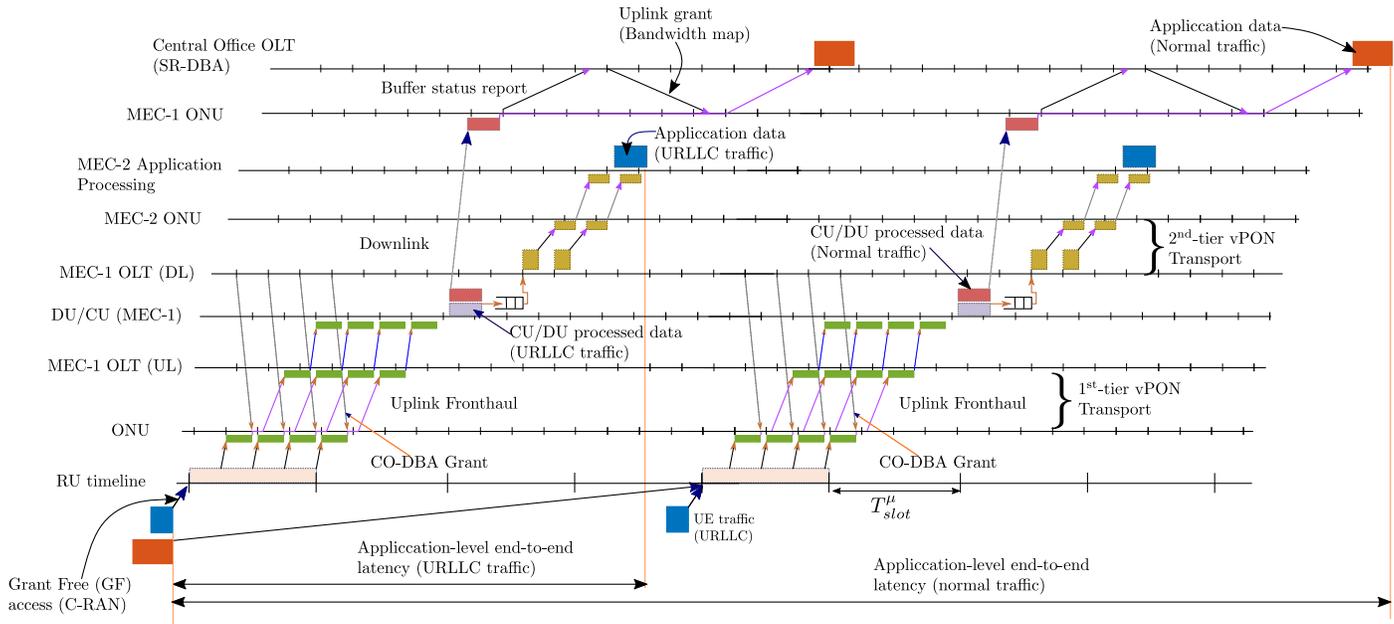}
			\caption{Two-tier vPON transport process illustrating the latencies involved in the various stages of the overall end-to-end transport}
			\label{fig:TwoTiervPONprotocol}
		\end{figure*}
	
		The second-tier vPON transport in our proposed architecture targets ultra-low latency for the connection between the DU/CU and the MEC nodes where the application is hosted (shown as MEC-2 in Fig. \ref{fig:SystemArc}). In a typical PON based fronthaul/midhaul/backhaul deployment, this connection between two MEC nodes is achieved by transporting the traffic via \ac{CO} which incurs a significant amount of latency. However, in our proposed architecture, this can be done by configuring the vPON slice of OLT-1 (at MEC-1) to temporarily include the ONU (that was originally connected with the control channel with CO \cite{MESH-JOCN}) at the MEC-2 and send the traffic to MEC-2 over the next downlink period of the same vPON slice. It is worth noting that in this work we assume a worst-case scenario, where the packet waits for the next downlink frame to start, although this could be further optimised in future work. The path of this CU/DU to application traffic via the proposed inter-MEC connectivity is illustrated with red-colored dashed line in Fig. \ref{fig:SystemArc}. In this case, an uplink transmission (in  the $1^{st}$-tier) is followed by a downlink transmission (in the $2^{nd}$-tier), thus the packet does not need to wait for a DBA scheduling round or implement a complex inter-PON Co-DBA coordination. 
		
		Fig. \ref{fig:TwoTiervPONprotocol} shows the whole two-tier vPON transport process and illustrates the latencies involved in the various location of the proposed architecture for achieving application-level low latency. As can be seen on Fig. \ref{fig:TwoTiervPONprotocol}, we consider two separate type of traffic: the URLLC traffic, which requires ultra-low latency of the order of 1-2ms, and the normal traffic that is generally latency tolerant (of the order of tens of milliseconds). The URLLC traffic uses the CGS PRB resources to transmit immediately at the next NR slot, while the normal traffic uses the standard RAN-MAC scheduling method, which requires about 4-NR slot times to acquire the PRB resources to transmit the data. 
		
		After the fronthaul reception and the CU/DU processing at the \ac{MEC}, the normal and URLLC UE traffic processed for that particular NR-slot are separated. At this point, the URLLC traffic is sent to the application hosted in the other MEC ($MEC_2$ in this case) and the normal UE traffic is sent to the application hosted at Cloud-Central office. This is sent uplink over the PON using the standard Status-Report based DBA (SR-DBA) mechanism. Therefore, the total application-level end-to-end latency mainly consists of the latencies incurred in the following interfaces of the overall connection path of this proposed architecture: UE access latency, Fronthaul transport latency ($1^{st}$-tier), queuing at CU/DU to MEC-1 OLT downlink interface, and CU-DU to application transport latency ($2^{nd}$-tier).
		
		The proposed two-tier vPON scheme can also be applied to the return path, where processed data is sent from MEC-2 to the UE. In this specific scenario, the return path consists of an uplink from MEC-2 to MEC-1, followed by a downlink from MEC-1 to the RU. The proposed scheme is effective in this case as well because, amount of the processed application response data for UE from MEC-2 is typically deterministic, therefore, a fixed bandwidth allocation with priority scheduling can be implemented to achieve low-latency at the uplink return path at the 2nd tier. This followed by the downlink path in the fronthaul to reach the response to the UE. It is worth noting that other configurations such as a two-tier downlink-downlink path, where the OLT at MEC-2 includes the ONU at MEC-1 in its vPON configuration to transmit traffic from MEC-2 to MEC-1 in a downlink PON, can also be explored. Investigating these alternative configurations presents interesting challenges related to load-balancing and maintaining ultra-low end-to-end application level latency.
					
		\item \textsl{\Large Control and Management}	
			
		One of the key operations of this architecture is the control and management of the slices, in order to assure coordination between the two-tiered transport. In our proposal, the configuration of vPON slices is facilitated by the OLT at the CO, which can send control information to both MECs using downlink PLOAM messages. As the ONUs at both MECs are initially connected to the control-channel with the CO OLT, the OLTs co-located at both MECs nodes receive the request for vPON slice configuration. In order to facilitate the $2^{nd}$-tier transport (which is downlink from MEC-1 OLT in this example), the ONU at MEC-2 tunes its wavelength to connect to OLT at MEC-1, while the OLT-1 updates its vPON slice to include the ONU at MEC-2 to complete the vPON slice reconfiguration process. At this point, any control channel information intended for the MEC-2 OLT is conveyed through the MEC-1 OLT (via downlink). Once this $2^{nd}$-tier transport is no longer required (i.e., the connection between CU/DU at MEC-1 application to MEC-2) possibly due to the application being migrated to some other MEC nodes, the ONU at MEC-2 goes back to control channel with the CO-OLT. The entire control and management information is conveyed through PLOAM messaging which is implemented in our system simulation.
	\end{enumerate}
	\section{Simulation Set Up} \label{sec:Simulation}
	The proposed architecture and the use case described above were simulated using OMNET++. The base architecture for the simulation setup follows a MESH-PON framework (for example as described in detail in \cite{MESH-Networks}), with fibre distances reported in Fig. \ref{fig:SystemArc}. The following enhancements were carried out on the MESH-PON simulator:
	\begin{enumerate}
		\item \textsl{\Large RU and user traffic}
		
		 On the wireless side, two user traffic arrival processes (normal and URLLC) are created following different Poisson processes. The \ac{CGS} is abstracted as a set of PRBs that are reserved for static pre-allocation whenever URLLC traffic arrives. Upon each user traffic arrival at RU (normal or URLLC), a group of PRB resources $N_{PRB}^{user}$ (which is chosen to be 5 in this work) is allocated. Here, each RU uses 4 MIMO antennas and a 7.2 split, operating over 100MHz of bandwidth, where a certain percentage of the PRBs (for example 10\% or 20\%) are semi-statically allocated/reserved as CGS resources that UEs can acquire for transmitting URLLC traffic immediately at the current 5G-NR slot. The rest of the PRBs are allocated using the standard PRB allocation process, following the resource request-and-grant process which takes about 4 NR slot-times. 
		
		\item \textsl{\Large Two-tier vPON Transport:}
		
		The transport architecture implements a MESH-PON with two-tier vPON transport, with the enhanced Co-DBA as proposed in Section \ref{sec:SystemModel} and the uplink-to-downlink switching scheme. In our simulation, we have incorporated a fiber propagation latency of 4.5 $\mu$s/Km. As a result, one-way transmission of traffic from MEC-MEC incurs a fixed latency of approximately $\approx$ 90 $\mu$s, while one-way transmission of traffic from MEC-Central office incurs a fixed latency of approximately $\approx$ 225 $\mu$s. In this work, we have taken into account the variability of DU/CU processing time at the MEC. To account for this type of latency, we have assumed that the processing time is proportional to the slot time. This is based on the assumption that shorter slot durations result in shorter timing windows for the CU/DU stack processing, as stated in \cite{DU_proc_Timing}. Therefore, we have used the slot time as a representation of the DU/CU processing time.). 
		
		We consider Split-7.2 for the fronthaul transport in this work. The fronthaul rate per RU-ONU with respect to the RU traffic load can be obtained from equation (\ref{eqn:fronthaul_rate_wrt_traff_load}). Here, $\delta_i$ is 1 if the corresponding PRB carries data traffic and 0 otherwise. A more in-depth explanation of this equation can be found in \cite{SplitPHY}.
		\begin{equation}
			\label{eqn:fronthaul_rate_wrt_traff_load}
			\begin{aligned}
				R_{\mathrm{7.2}}=&\left(R_{\mathrm{PUSCH}, \mathrm{IQ}}+R_{\mathrm{DMRS}, \mathrm{IQ}}\right)+R_{\mathrm{PUCCH}, \mathrm{IQ}} \\
				 &+R_{\mathrm{PRACH}, \mathrm{IQ}} + R_{\mathrm{SRS}, \mathrm{IQ}} \\
				=&2 N_{\mathrm{ant}}\left(\sum_{i=1}^{N_{\mathrm{RB}}} N_{\text {res }, i} N_{\mathrm{RE}, i} \delta_i \frac{1}{T_{slot}^{\mu}}\right. \\
				 &\quad    +N_{\text{reg }}^{\text{PUCCH }} N_{\mathrm{RE}}^{\mathrm{PUCCH}} N_{\text {res}}^{\text {PUCCH }} \frac{1}{T_{slot}^{\mu}}\\
				 &\quad\qquad    + N_{\text {bins }}^{\text {PRACH}} N_{\text {res, PRACH }} \frac{1}{T_{\text {PRACH }}} \\
				 &\quad\qquad\qquad    \left.+N_{\text {scrr,SRS }} N_{\text {res, SRS }} \frac{1}{T_{\mathrm{SRS}}}\right)
			\end{aligned}
		\end{equation}
		\begin{table}[t]
			\centering
			\caption{Simulation parameters for RU and user traffic}
			\label{tab:paramsSIM}
			\begin{tabular}{lc|c}
				\hline
				\multirow{2}{*}{Parameters} & \multicolumn{2}{c}{values} \\ \cline{2-3}
				& Config-1 & Config-2\\ \hline
				NR Bandwidth (per RU) & \multicolumn{2}{c}{100 MHz} \\
				NR numerology ($\mu$) & 1 & 2\\ 
				NR Slot Time ($T_{slot}^{\mu}$) & 0.5 ms & 0.25 ms \\
				maximum No. of PRBs ($N_{PRB}^{BW(j)}$) & 270 & 135\\
				Num PRBs per user ($N_{PRB}^{user}$) & \multicolumn{2}{c}{5} \\
				Traffic type & \multicolumn{2}{c}{\{Normal , URLLC\}} \\
				\begin{tabular}{@{}l@{}}
					Percentage of PRBs reserved \vspace{-0.2cm} \\
					for CGS (URLLC traffic)
				\end{tabular} & \multicolumn{2}{c}{\{10\% , 20\%\}} \\
				Num MIMO layers per RU ($v^{(j)}_{Layers}$) & \multicolumn{2}{c}{4} \\
				Num antennas per RU ($N_{\mathrm{ant}}$) & \multicolumn{2}{c}{4} \\
				Modulation order ($Q^{(j)}_{m}$) & \multicolumn{2}{c}{256 QAM} \\
				\begin{tabular}{@{}l@{}}Number of component carriers \vspace{-0.2cm} \\ for carrier aggregation ($J$)\end{tabular} & \multicolumn{2}{c}{1} \\
				Scaling Factor ($f^{(j)}$) & \multicolumn{2}{c}{1}\\
				$R_{max}$ & \multicolumn{2}{c}{$948/1024$} \\
				Overhead ($OH^{(j)}$) & \multicolumn{2}{c}{\begin{tabular}{@{}c@{}} 0.1 (for Frequency\vspace{-0.2cm} \\ range FR1 and UL) \end{tabular}} \\
				\begin{tabular}{@{}c@{}}OFDM symbol duration\vspace{-0.2cm} \\including CP (in $\mu$s)\vspace{-0.15cm}\\ \big($T_s^\mu = 10^{-3}/(14*2^{\mu})$\big) \end{tabular} & 35.714 $\mu$s & 17.85 $\mu$s\\
				\hline
				
			\end{tabular}
		\end{table}
		
		\begin{table}[t]
			\centering
			\caption{Simulation parameters for two-tier vPON transport}
			\label{tab:paramsPON}
			\begin{tabular}{lc}
				\hline
				Parameters & Values\\ \hline
				\begin{tabular}{@{}l@{}}
					(TWDM-PON) Uplink capacity per\\
					OLT channel (CO-OLT, MEC-OLT)
				\end{tabular}
				 & 50 Gbps \\
				Fiber Propagation delay & 4.5 $\mu$s/Km \\
				OLT grant cycle (GC) duration & 125 $\mu$s \\
				Ethernet (eCPRI) frame size & 2048 bytes \\
				Inter packet gap for eCPRI packets & $10^{-7}$s\\
				ONU response time & 35 $\mu$s \\
				CU/DU stack processing delay & 0.5ms , 0.25ms \\ \hline
				\begin{tabular}{@{}l@{}}
					eCPRI 7.2 rate \vspace{-0.15cm} \\
					calculation parameters
				\end{tabular} &  
				\hspace{-0.5cm}\begin{tabular}{@{}c@{}}
					$N_{\mathrm{RE}, i}$=156,$N_{\text {res }, i}$=8 \\
					$N_{\text{reg }}^{\text{PUCCH }}$ = 1\\
					$N_{\mathrm{RE}}^{\mathrm{PUCCH}}$= 156 \\
					$N_{\text {res}}^{\text {PUCCH}}$=8\\
					$N_{\text {bins }}^{\text {PRACH}}$ = 839\\
					$N_\text{res,PRACH}=10$ \\
					$T_{\text {PRACH }}=$ 10 ms\\
					$N_{\text {RE, SRS }}$=12\\
					$N_{\text {res, SRS }}$=8 \\
					$N_{\text {scrr,SRS }}=((N_{PRB}^{BW(j)}-$\\
					$N_{\text{reg }}^{\text{PUCCH }})*N_{\text {RE, SRS }})$ \\
					$T_{SRS}=$1 ms \\
				\end{tabular}\\
				\hline				
			\end{tabular}
		\end{table}
		
		Using equation (\ref{eqn:fronthaul_rate_wrt_traff_load}), at the end of each NR-slot, the user traffic at the RU is converted to fronthaul traffic depending on how many users are currently active on the current NR slot (consequently the number of PRBs carrying data). Therefore, given a vPON slice configuration (i.e, the number of RU-ONUs in the vPON slice), we can use this to calculate the PON load (or traffic intensity) as follows:\\
		$\displaystyle \text{Traffic Intensity (\%load)} = \frac{\sum_{i=1}^{N_{sl}}R^{i}_{7.2}}{T_{cap}}\times 100$\\
		Where $N_{sl}$ represents the number of RU-ONUs in the vPON slice, $R^{i}_{7.2}$ denotes the fronthaul rate for $i^{th}$ RU in the vPON slice obtained using (\ref{eqn:fronthaul_rate_wrt_traff_load}), and $T_{cap}$ represents the capacity per OLT channel (which is considered to be 50 Gbps in this case).

		After the fronthaul reception and the CU/DU processing at \ac{MEC}, the normal and URLLC UE traffic processed for that particular NR-slot is separated. At this point, the URLLC traffic is sent to the application hosted in the other MEC ($MEC_2$ in this case) and the normal UE traffic is sent to the Cloud-Central office for further application processing of the normal traffic. The amount of DU processed data per NR-Slot that is to be sent to the application processing largely depends on the amount of each traffic carried on that NR-Slot (i.e, the number of PRB resource occupied). For example, for the URLLC traffic, the amount of DU processed data for a particular NR-Slot $D_{du}^i$ to be sent from $MEC_1$ to the application processing at $MEC_2$ depends on the number of PRBs occupied on the corresponding NR-slot, and is calculated using the following equation (\ref{eqn:data_size_per_slot}).
		\begin{equation} \label{eqn:data_size_per_slot}
			D_{du}^i \text{ (in Mb)} = \Bigl( \bigl(R_{cell}/N_{PRB}^{BW(j),\mu}\bigr) \,.\, N_{PRB}^{user} \,.\, u_{slot}^{i} \,.\, T_{slot}^{\mu}\Bigr)
		\end{equation}
		Here, $N_{PRB}^{user}$ denotes the number of PRB resources allocated per user traffic instance. $u_{slot}^{i}$ and $T_{slot}^\mu$ denote the number of active users in the slot and NR-slot duration, respectively. $R_{cell}$ denotes the maximum cell throughput calculated using equation (\ref{eqn:cell_throughput}) and is based on 3GPP TS 38.306 \cite{3gppts38.306}. 
		\begin{equation} \label{eqn:cell_throughput}
			\begin{aligned}
					R_{cell} \text{ (in Mbps)} \,=\, & 10^{-6}.\sum_{j=1}^{J} \Biggl( v^{(j)}_{Layers} . Q^{(j)}_{m} . f^{(j)} . R_{max} . \Bigr. \\
				  & \vspace{2cm} \qquad \frac{N_{PRB}^{BW(j),\mu} . 12}{T_s^\mu} \bigl. . \left( 1 - OH^{(j)}\right) \Biggr)				
			\end{aligned}
		\end{equation}
		Tables \ref{tab:paramsSIM} and \ref{tab:paramsPON} reports the parameters used for the simulation and for generating the results.
	\end{enumerate}


	\section{Results} \label{sec:Results}
	We run our simulations repeatedly for 30 second intervals (which is about 240,000 OLT grant cycles). We collected latency metrics for each received packets at various location of the proposed architecture, counting between 4 million and 3 billions values, depending on the observation location and the traffic load.
	
\begin{figure}[h]
		\centering
		\includegraphics[clip, trim={0 0 0 0}, width=0.75\linewidth]{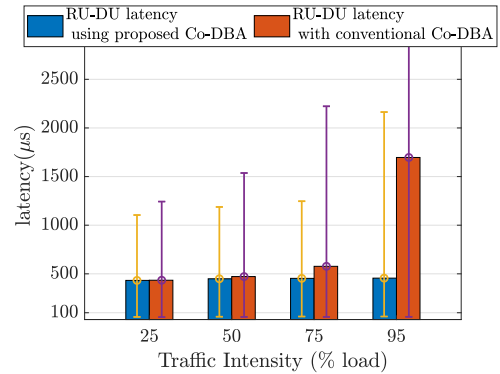}
		\caption{Latency performance comparison of proposed enhanced Co-DBA vs the conventional Co-DBA.}
		\label{Fig:prop_Co-DBA_vs_conv_Co-DBA}
	\end{figure}

	Fig. \ref{Fig:prop_Co-DBA_vs_conv_Co-DBA} shows end-to-end latency in the first tier of the vPON transport i.e., fronthaul latency between the RU and DU. It also demonstrates the advantage of our enhanced Co-DBA over the conventional Co-DBA. Here, the conventional DBA, which does not take into account CGS allocated resources, allocates bandwidth based on the scheduling information of normal-UE traffic (4-NR slot prior) and adds a fixed-bandwidth corresponding to 5\% of the overall RU traffic to accommodate for fluctuations due to URLLC traffic. Our proposed Co-DBA, however, takes into account both the allocation of normal-UE scheduling information (4-NR slot prior) and the allocation of semi-static CGS resources obtained from RRC and passed through the CTI interface to the OLT, resulting in significant improvements in latency, particularly at higher loads where URLLC-traffic at CGS resources increases significantly and improper DBA allocation by conventional DBA causes increased queuing latency at the fronthaul between RU and DU.
	
\begin{figure}[h]
		\centering
		\includegraphics[clip, trim={0 0 0 0}, width=\linewidth]{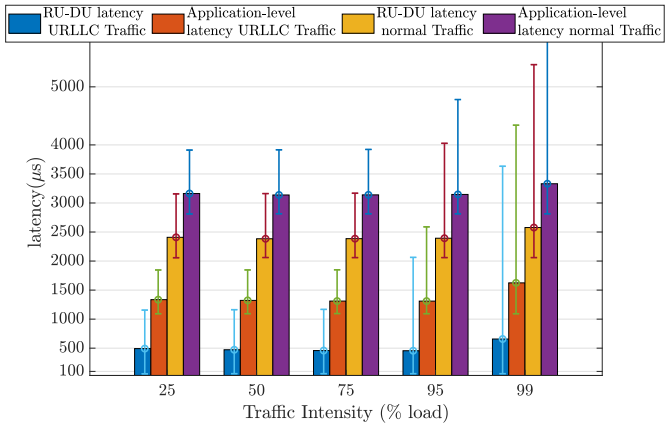}
		\caption{Average and Max application-level end-to-end latency for different traffic types and PON load (slot time =0.5ms).}
		\label{Fig:tow-tiervPON}
	\end{figure}

	Fig. \ref{Fig:tow-tiervPON} shows the end-to-end latency, measured both at the RU-DU interface and at the application level (i.e., end-to-end), of our proposed mechanism, against the traffic load on the PON uplink. This figure also illustrates the end-to-end latency difference between the URLLC traffic (where we employ the two-tier vPON scheme) and the normal traffic (which uses ordinary RAN scheduling, Co-DBA and a main OLT at a CO that is 50 km away). The normal traffic uses PON fronthaul followed by conventional SR-DBA to transport CU/DU traffic to the application at the CO, and serves as the benchmark in this study. As can be seen, our proposed scheme can achieve end-to-end (application-level) average latency just above 1 ms (red bar), with maximum latency around 1.7 ms. This remains unaffected by load until around 95\% traffic, when the average latency increases above 2.5ms (although the average remains approximately the same).  It is important to note that these results do not take into account the latency generated by the application processing, as that would depend on the specific application being run. The results only show the latency of the communication between the applications.
	
\begin{figure}[h]
		\centering
		\includegraphics[clip, trim={0 0 0 0}, width=\linewidth]{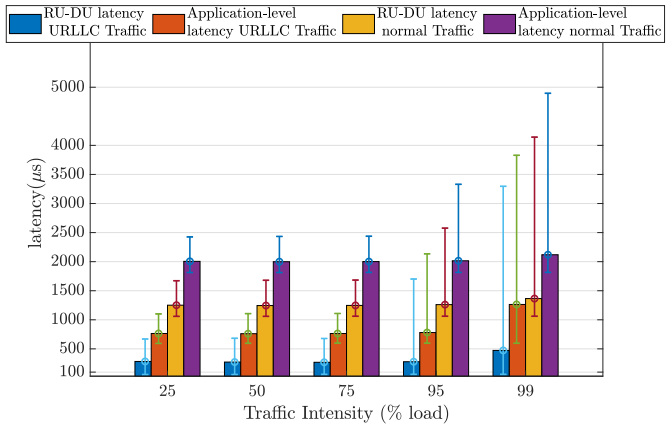}
		\caption{Average and Max application-level end-to-end latency for different traffic types and PON load (slot time =0.25ms).}
		\label{Fig:tow-tiervPON_Slot0_25ms}
	\end{figure}

\begin{figure}[h]
		\centering
		\includegraphics[clip, trim={0 0 0 0}, width=0.9\linewidth]{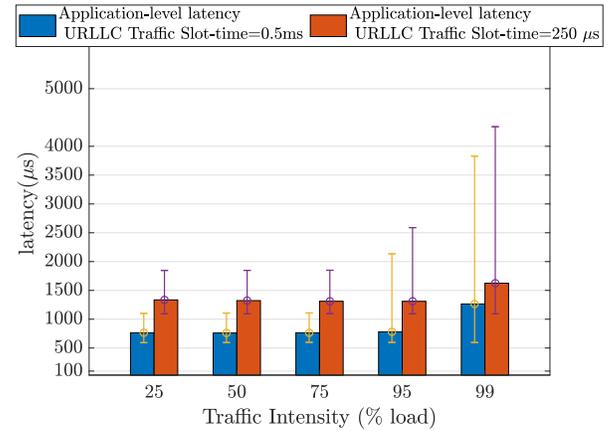}
		\caption{Comparison of Average and Max application-level end-to-end latency at different traffic load for time slot of 0.25ms and 0.5ms.}
		\label{Fig:tow-tiervPON_diffSolt}
	\end{figure}

	A way to further reduce latency is to reduce the RAN slot duration from 0.5 ms (used for fig. \ref{Fig:tow-tiervPON}), to 0.25 ms. This is shown in Fig. \ref{Fig:tow-tiervPON_Slot0_25ms} and \ref{Fig:tow-tiervPON_diffSolt}. 
	Using a shorter NR slot configurations we are able to meet sub millisecond latency (both average and max) up to about 95\% load. This is because, shorter time slot reduces the waiting time for the user traffic. This also reduces the CU/DU processing time window. Therefore, the overall application-level latency is reduced.
	
	
\begin{figure}[H]
		\centering
		\includegraphics[clip, trim={0 0 0 0}, width=\linewidth]{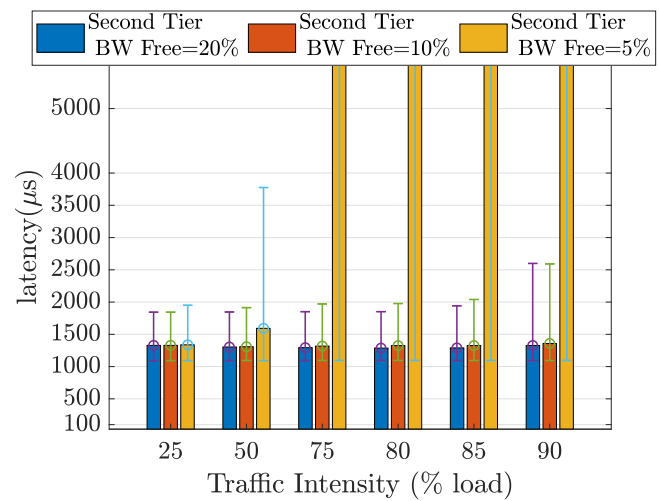}
		\caption{Average end-to-end application-level latency at different traffic load for different fraction of the available OLT bandwidth (20\% 10\% and 5\%) for $2^{nd}$-tier transport (slot-time=0.5ms).}
		\label{Fig:tow-tiervPON_diffBWfreeSlot0_5ms}
	\end{figure}

\begin{figure}[h]
		\centering
		\includegraphics[clip, trim={0 0 0 0}, width=\linewidth]{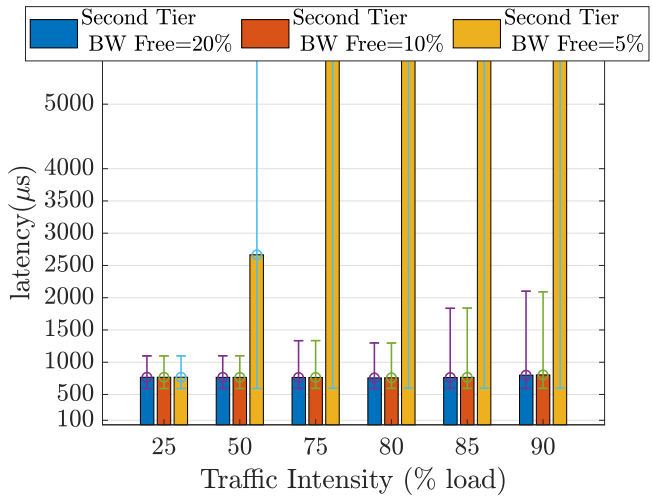}
		\caption{Average end-to-end application-level latency at different traffic load for different fraction of the available OLT bandwidth (20\% 10\% and 5\%) for $2^{nd}$-tier transport (slot-time=0.25ms).}
		\label{Fig:tow-tiervPON_diffBWfreeSlot0_25ms}
	\end{figure}

	In both the results above, we have considered that sufficient downlink bandwidth is always available at the edge-OLT (at the MEC location) for URLLC traffic from the DU to the application located at the other MEC, via the second-tier vPON. However, in practice, only a fraction of the total OLT downlink bandwidth may be available, since the OLT also serves downlink fronthaul for other RUs in that vPON. This fraction of available bandwidth would depend on various factors, for example the downlink traffic load per RU, the functional split and the number of RUs in the vPON slice. Fig. \ref{Fig:tow-tiervPON_diffBWfreeSlot0_5ms} and \ref{Fig:tow-tiervPON_diffBWfreeSlot0_25ms} shows the overall application-level latency when we reduce the available downlink bandwidth for the second tier vPON transport. We report the analysis for value of available bandwidth between 25\% and 5\% (i.e., 75\% to 95\% of the OLT downlink bandwidth respectively is occupied by other fronthaul services). Here, we consider taht 20\% of the overall traffic is URLLC (i.e, having ultra-low latency requirement). Therefore, an increase in overall RU traffic load also means a proportional increase of URLLC traffic. As a consequence, this also increases the traffic at the second-tier path between CU/DU and application at MEC-2. Therefore, given only a fraction of the OLT downlink bandwidth is available for the second-tier transport, it is important to analyse how much traffic load at the RU can be supported with the required ultra-low application-level end-to-end latency.
	

	As it can be seen from this figures, our proposed scheme can achieve maximum end-to-end latency below 2ms (for 0.5ms NR-slot) or $\approx$1ms (for 0.25ms NR slot) for traffic load up to about 80\%, when at least 10\% downlink bandwidth is available to the OLT for 2nd-tier PON transport of DU processed traffic towards the application. However, as this available bandwidth reduces down to 5\%, then less than 50\% of the traffic load can achieve ultra-low end-to-end latency. Above this load, although the fronthaul latency is still low ($\approx$ 450$\mu$s) the queuing latency at the interface between DU and OLT downlink interface (for 2nd-tier vPON transport to application) increases and therefore, the overall latency is over 5 ms.


	\vspace{-2mm}
	\section{Conclusions} \label{sec:Conclusion}
	In this paper, we have proposed a novel two-tier PON transport method with schedulers coordination over a virtualised MESH-PON to achieve application-level ultra-low latency. Our proposed method addressed three major sources of latency, namely: RAN access latency, Fronthaul latency and the CU/DU to application transport latency. While the RAN access is significantly reduced by using CGS resources to transport URLLC traffic, we have proposed a modification of the Co-DBA to incorporate the traffic at the CGS resources while calculating the uplink grants for maintaining low fronthaul latency. To address the CU/DU to application path latency (backhaul), we have proposed a second-tier vPON transport to enable direct communication between MEC-nodes without layer 2 switching back at the CO OLT, thus reducing latency significantly. The results show application-level end-to-end transport latency of the order of 2 ms, depending upon fronthaul and RU traffic load when using 0.5 ms NR slot. A further reduction in the end-to-end latency $\approx$1ms can be achieved by using a even shorter NR slot duration (for e.g, 0.25 ms). Therefore, in conclusion, we have shown how the use of a virtualised MESH-PON with coordinated scheduling can be instrumental in supporting URLLC latency requirements of the order of 1 ms, on a network topology covering distances up to 20 km (while ordinary traffic can be served by OLTs in the CO located more than 50 km away). The use of MESH-PON is key to support low-cost connectivity to enable densification of small cells and MEC nodes, which is a key factor for delivering the high capacity, reliability and coverage required by beyond 5G networks and applications.

	%
	
	%
	%
	%

	\section*{Acknowledgments}
	Financial support from SFI grants 17/CDA/4760, 18/RI/5721 and 13/RC/2077 is gratefully acknowledged.

	\bigskip

\begin{thebibliography}{99}
		\bibitem{5G-NGMN-Verticals} "Verticals URLLC Use Cases and Requirements". NGMN Allaince, Feb 2020.
		
		\bibitem{5G-NR_Dahlman-Scheduling} E. Dahlman, S. Parkvall, and J. Sk\"{o}ld. "5G NR: the next generation wireless access technology, CHAPTER 14: Shceduling". Elsevier, Academic Press, 2021.
		
		\bibitem{5G-NR_Dahlman} E. Dahlman, S. Parkvall, and J. Sk\"{o}ld. "5G NR: the next generation wireless access technology, CHAPTER 20: Industrial IoT and URLLC Enhancements". Elsevier, Academic Press, 2021.
		
		\bibitem{ORAN_FH_Standard} "O-RAN Fronthaul Control, User and Synchronization Plane Specification 8.0." In Specification WG4: Open Fronthaul Interfaces Workgroup, Mar 2022.
		
		\bibitem{Co-DBA} T. Tashiro, S. Kuwano, J. Terada, T. Kawamura, N. Tanaka, S. Shigematsu, and N. Yoshimoto. "A novel DBA scheme for TDM-PON based mobile fronthaul". Tu3F.3, OFC 2014.
		
		\bibitem{O-RAN-CTI} O-RAN Fronthaul Working Group 4. "Cooperative Transport Interface Transport Control Plane Specification". O-RAN alliance, Mar. 2021.
		
		\bibitem{vDBA} M. Ruffini, A. Ahmadi, S. Zeb, N. Afraz and F. Slyne. "The Virtual DBA: Virtualizing Passive Optical Networks to Enable Multi-Service Operation in True Multi-Tenant Environments". OSA Journal of Optical Communications and Networking, No.4, Vol.12, April 2020.
		
		\bibitem{MESH-Networks} S. Das, F. Slyne and M. Ruffini. "Optimal Slicing of Virtualised Passive Optical Networks to Support Dense Deployment of Cloud-RAN and Multi-Access Edge Computing". IEEE Network Vol.36,No.2, March/April 2022.
		
		\bibitem{MESH-JOCN} S. Das, F. Slyne, A. Kaszubowska and M. Ruffini. "Virtualised EAST-WEST PON Architecture Supporting Low-Latency communication for Mobile Functional-Split Based on Multi-Access Edge Computing". JOCN, 10(12), Oct. 2020
		
		\bibitem{Wong} C. Ranaweera, E. Wong, C. Lim, and A. Nirmalathas. "Next generation optical-wireless converged network architectures". IEEE Network 26, 22-27 (2012).
		
		\bibitem{Pfeiffer} T. Pfeiffer. "Converged heterogeneous optical metro-access networks". Tu.5.B.1, OFC 2010.
		
		\bibitem{ECOC2022} S Das, F Slyne, D Kilper and M Ruffini. "Schedulers Synchronization Supporting Ultra Reliable Low Latency Communications (URLLC) in Cloud-RAN over Virtualised Mesh PON", European Conference in Optical Communications (ECOC), Sept 2022.
		
		\bibitem{C-RAN-multitenancy} X. Li et al., "5G-Crosshaul Network Slicing: Enabling Multi-Tenancy in Mobile Transport Networks," in IEEE Communications Magazine, vol. 55, no. 8, pp. 128-137, Aug. 2017, doi: 10.1109/MCOM.2017.1600921.
		
		\bibitem{DU_proc_Timing} J. S. Panchal, S. Subramanian and R. Cavatur, "Enabling and Scaling of URLLC Verticals on 5G vRAN Running on COTS Hardware", in IEEE Communications Magazine, vol. 59, no. 9, pp. 105-111, September 2021
		
		\bibitem{SplitPHY} U. Dotsch, M. Doll, H.-P. Mayer, F. Schaich, J. Segel, and P. Sehier, "Quantitative analysis of split base station processing and determination of advantageous architectures for LTE," Bell Labs Technical Journal, vol. 18, no. 1, pp. 105–128, Jun. 2013, doi: 10.1002/bltj.21595.
		
		\bibitem{3gppts38.306} 3GPP TS 38.306, "5G; NR; User Equipment (UE) radio access capabilities," version 16.10.0 Release 16. \url{https://www.etsi.org/deliver/etsi_ts/138300_138399/138306/16.10.00_60/ts_138306v161000p.pdf}
		
	\end{thebibliography}
\end{document}